\documentclass[prb,aps]{revtex4}
\usepackage{epsfig,latexsym,amssymb,amsmath,amsbsy,graphics,graphicx}

\begin{document}

\title{Large Moment Formation and Thermodynamic Properties of
Disordered Spin Ladders with Site Dilution}

\author{Eddy Yusuf and Kun Yang}
\affiliation{
National High Magnetic Field Laboratory and Department of Physics,
Florida State University, Tallahassee, Florida 32306
}

\date{\today}

\begin{abstract}
Low temperature properties of antiferromagnetic two-leg spin-1/2 ladders with 
bond randomness and site dilution (or doping with nonmagnetic impurities)
are studied using the real-space 
renormalization-group technique. We find that for non zero dopant
concentrations the systems are driven into a phase dominated by large 
effective spins, i.e. the Large Spin phase. The susceptibility follows a 
universal Curie-like $1/T$ behavior at low temperature,
regardless of the dopant concentration (as long as it is nonzero)
and the strength of bond
randomness. Very similar behavior has been found in ladders that are doped
with magnetic impurities that carry spin-1.
\end{abstract}

\maketitle

\section{Introduction}

Quantum effects in one-dimensional spin systems have 
attracted prolonged interest from 
both theoretical and experimental physicists. These include quasi-long 
range order, topological order, 
and fluctuation induced excitation gap ({\em e.g.}, the Haldane gap) that  
are purely quantum mechanical effects enhanced by
the low-dimensionality of the systems. 
Among these quantum phenomena, the effects of disorder have been studied 
by many groups. It was found that disorder can 
qualitatively change the low temperature physics and produce
rich disorder-dominated phases in these systems. 
One class of such systems which have received considerable 
attention are random antiferromagnetic spin chains. Most of the theoretical 
studies 
of random spin chains are based on the real space
renormalization group (RSRG) 
method developed by Ma, Dasgupta, and Hu \cite{mdh} in 
this context, 
and Bhatt and Lee 
\cite{bl} in the study of doped semiconductors. This RSRG method was extended
further by Fisher \cite{fisher1} and allows one to obtain results which are 
essentially exact for the random 
spin-1/2 chain. The application of this method to
the other random spin chain models by a number of authors \cite{fisher2,
westerberg,hybg,boechat,hy,monthus,yb,damle} has given us a better 
understanding of the behavior of these systems at low temperature.

Another example of 1D spin system that is of considerable recent interest
is the two-leg
antiferromagnetic spin-1/2 ladder.\cite{dagotto} It is known to have
an excitation gap similar to the Haldane gap of integer AF spin chains, and
short-range spin-spin correlation.
Compared to the spin chains, only relatively few theoretical studies have been
devoted to the study of disorder effects in spin ladders. 
Several authors have investigated
the effects of bond randomness.\cite{og,melin,ek} 
It was found that the ladder is remarkably stable against {\em weak} 
bond randomness.\cite{og} Stronger randomness introduces  
a large density of low-energy excitations into the system,\cite{melin,ek}
which can lead to {\em divergent} spin susceptibility in the limit 
$T\rightarrow 0$.\cite{ek} However, the spin-spin correlation remains 
short-ranged,\cite{ek} contrary to what occurs in strongly disordered
AF spin chains.\cite{fisher1,hy}

In real systems, bond randomness is typically induced by impurities
{\em away} from the ladder, which distort the lattice structure (and hence
coupling constants) without affecting the spins that form the ladder. 
Experimentally, another way to introduce and control disorder in the system is
to introduce dopants that go directly into the ladder, so that the ions that
carry the half-spin (typically the Cu ion) are randomly
replaced by nonmagnetic ions
(like Zn), or ions with other spin sizes (like Ni which carries spin-1).
Such disorder not only induces lattice distortion, but also changes the
lattice structure of the spin ladder through site-dilution etc, and thus has
more dramatic effects.
A number of experimental\cite{azuma,fujiwara} and
theoretical\cite{motome,sigrist,iino,martins2,miyazaki,sandvik,martins,
mikeska,laukamp,steiner,gogolin} works have been 
devoted to study the doped two-leg spin-1/2
ladder, for example Sr(Cu$_{1-x}$Zn$_x$)$_2$O$_3$, where 
some Cu ions are replaced by 
nonmagnetic Zn ions. It was found experimentally 
that even a small amount of nonmagnetic 
doping is enough to change the low-temperature behavior of the
systems drastically, and gives rise to divergent susceptibility at low 
temperature. Theoretically, it is understood that
a {\em single} Zn impurity induces an effective, localized spin-1/2 moment 
in the vicinity of the dopant; such localized moments immediately destroy 
the spin gap. \cite{motome,martins2,martins}
When there is a small but finite density of dopants, these effective 
spin-1/2 moments interact with each other, and currently there is no consensus
on
what the asymptotic low-temperature behavior is.
Sigrist and Furusaki\cite{sigrist} argued that the system can be mapped onto
an effective model that is made of these 
{\em effective} half spins induced by the
dopants forming a spin-1/2 chain, with random antiferromagnetic (AF)
and ferromagnetic (F)
couplings; this model is known to form large effective spins and exhibits Curie
susceptibility at low $T$: $\chi\sim 1/T$.\cite{westerberg} 
On the other hand, Gogolin {\em et al.}
\cite{steiner,gogolin} used the bosonization method to map the 
problem to a Dirac fermion with random mass, and concluded that 
the low $T$ susceptibility behaves as
$\chi\sim 1/(T\log^2T)$, which is the same as the random singlet
phase;\cite{fisher1} no large moment formation was found in their work.
Existing exact diagonalization\cite{motome} and
quantum Monte Carlo calculation\cite{motome,iino,miyazaki} 
do not have large enough
system size to unambiguously resolve this discrepancy.

In this work we study disordered two-leg spin ladders with both bond randomness
and site dilution (corresponding to Zn doping), using the RSRG method.
As discussed earlier, in principle Zn doping introduces two types of disorder.
Technically the presence of bond randomness is useful 
to us in our study as it introduces
a separation of energy scales and justifies the usage of the RSRG method.
Using this method we are able to study systems with sizes 100 times larger
than those accessible in quantum Monte Carlo studies.
In addition to the 
nonmagnetic (Zn) doping, we also study theoretically
for the first time magnetic doping by
replacing the Cu ions with Ni ions (or doping with $S=1$ impurities),
a situation already realized experimentally.\cite{ohsugi}

Our results are summarized as follows. As the
RSRG procedure is carried out, effective spins (or moments) larger than 1/2
start to form;
these large moments persist and grow 
without bound as the energy scale is lowered,
regardless of the dopant concentrations (as long as it is non-zero)
or the strength of bond
randomness. Thus the presence of dopants drives the system into a new phase 
which is controlled by large spins, i.e. 
the Large Spin phase; the susceptibility at low
temperature remains universal and follows $1/T$ behavior as $T$ goes to zero. 
The $1/T$ Curie behavior comes from the spins coupled together forming larger 
effective spins. 
Such behavior is very similar to that of random AF-F
spin chain studied by Westerberg {\em et al.},\cite{westerberg} 
as anticipated by Sigrist and
Furusaki. While for any finite temperature range it is very difficult to 
distinguish between $\chi\sim 1/T$ and $\chi\sim 1/(T\log^2T)$, we further 
find that the Curie coefficient of $\chi$ approaches that predicted by 
Sigrist and Furusaki based on analogy to the random AF-F spin chain.
We thus conclude that the low energy behaviors of disorder spin ladders are
the same as those of random AF-F spin chains. 

The remainder of the paper is organized as follows. In Sec. II we introduce 
the model we use and review the application of RSRG to this model. In Sec.
III we present our numerical results, compare them to previous works, and 
discuss the significance of our results. In Sec. IV we 
summarize our work and discuss the implications of our results.

\section{The Model and Review of previous results}

The model we consider in this work describes a disordered
antiferromagnetic two-leg 
spin-1/2 ladder. The Hamiltonian for this model is given by :
\begin{equation}
\label{ladder}
H = \sum_{i=1}^{N-1}\sum_{j=1,2} J_{i,j} \boldsymbol{S}_{i,j}\cdot\boldsymbol
{S}_{i+1,j} + \sum_{i=1}^{N} K_{i} \boldsymbol{S}_{i,1}\cdot\boldsymbol
{S}_{i,2},
\end{equation}
where $\boldsymbol
{S}_{i,j}$ is a spin-1/2 operator when there are no dopants, 
and the positive coupling constants 
$J_{i,j}$ (couplings along the chains, or legs of the ladder) and $K_i$ 
(couplings between the chains, or along the rungs of the ladder) are 
distributed randomly according to some probability 
distributions $P_{\parallel}(J_{i,j})$ and $P_{\perp}(K_i)$.
$N$ is used to represent the number of sites for a single chain.
The dopant concentration is given by $z$, namely we put $2Nz$ nonmagnetic
impurities (Zn doping) or  magnetic impurities (Ni doping) on the ladder;
for nonmagnetic impurities we simply remove the spins at the impurity sites,
while for magnetic impurities we replace the spin-1/2 operators by 
spin-1 operators at the impurity sites.

We use the real-space renormalization-group 
(RSRG) method to study this problem. 
Application of the RSRG procedure with proper extensions to the ladder
systems has been discussed at length in Ref. \onlinecite{ek}, 
and we refer the readers to 
that article for details. For the purpose of later comparison, 
here we briefly review some of the relevant results of that work, which
studies the effects of bond randomness {\em without} non-magnetic or magnetic
(spin-1) dopants going into the ladder.

When the RSRG procedure is carried out in ladders with no dopants, it was found
that ferromagnetic (F) couplings are generated, and some
effective spins with sizes
bigger than 1/2 are formed because these ferromagnetic bonds may become the 
strongest
bond in the system at some stage of RG.\cite{ek} However, the
percentage of these large effective spins remains low at all stages of RG, and 
in the low-energy limit, their percentage {\em decreases} as the energy scale is
going down, due to the fact that the overall strength of the ferromagnetic 
bonds becomes much
weaker than that of the antiferromagnetic bonds, even though they have 
roughly the
same numbers. Such behavior may be understood in the following way.
With nearest neighbor couplings only, the ladder has a bipartite lattice 
structure which means the system can be divided into two sublattices (A and B),
and spins
sitting on sublattice A are always coupled to spins sitting on sublattice B,
and vice versa. In the absence of dopants, the number of spins in the two
sublattices are strictly equal, and Marshall's theorem \cite{marshall}
dictates that the ground state is a total spin singlet in this case. 
Heuristically this is easy to see: the spins in the same sublattice tend to be
parallel while those in opposite sublattices tend to be antiparallel, and there
is a total cancellation when the number of spins are the same in the two 
sublattices. The disappearance of large effective spins in the low-energy limit
is simply a reflection of this cancellation effect.

As we will see below, the situation becomes very different in the presence of
dopants. In this case the dopants go onto lattice sites randomly, thus there 
are
{\em fluctuations} in the numbers of dopants going onto the two different 
sublattices, even though {\em on average} they are the same. Such fluctuations
destroy the perfect cancellation discussed above, and as we see below, lead to
large moment formation in the long-distance, low-energy limit, which in turn
changes the thermodynamic properties of the system qualitatively.

\section{Numerical Results}

We present numerical results for the spin ladder with the length 
of the ladder up to 100 000 (200 000 total spins). 
We decimate the strongest bond in the system defined as the bond with the
largest energy gap, $\Delta_0$, between the ground state and the first 
excited state. The decimation process is repeated
until the number of spins left is less than 1\% of the original number of 
spins in the system. The initial distributions are taken 
to be in power-law form:\cite{note}
\begin{eqnarray}
\label{distribution}
P_{\parallel}(J_{i,j}) &=& (1-\alpha) J_{i,j}^{-\alpha}, 0 < J_{i,j} < 1;
\nonumber\\
P_{\perp}(K_i) &=& \frac{1-\alpha}{\Lambda^{1-\alpha}} K_{i}^{-\alpha},
0 < K_i < \Lambda.
\end{eqnarray}
Here $0 \leq \alpha < 1$ is the measure of bond disorder (the bigger $\alpha$,
 the
stronger the randomness strength), and $\Lambda$ is the anisotropy 
parameter; in the limit $\Lambda\rightarrow 0$ the two chains decouple. The
nonmagnetic or $S=1$ magnetic dopants are distributed
randomly throughout the system. 

\begin{figure*}
\includegraphics*[angle=-90,scale=0.6]{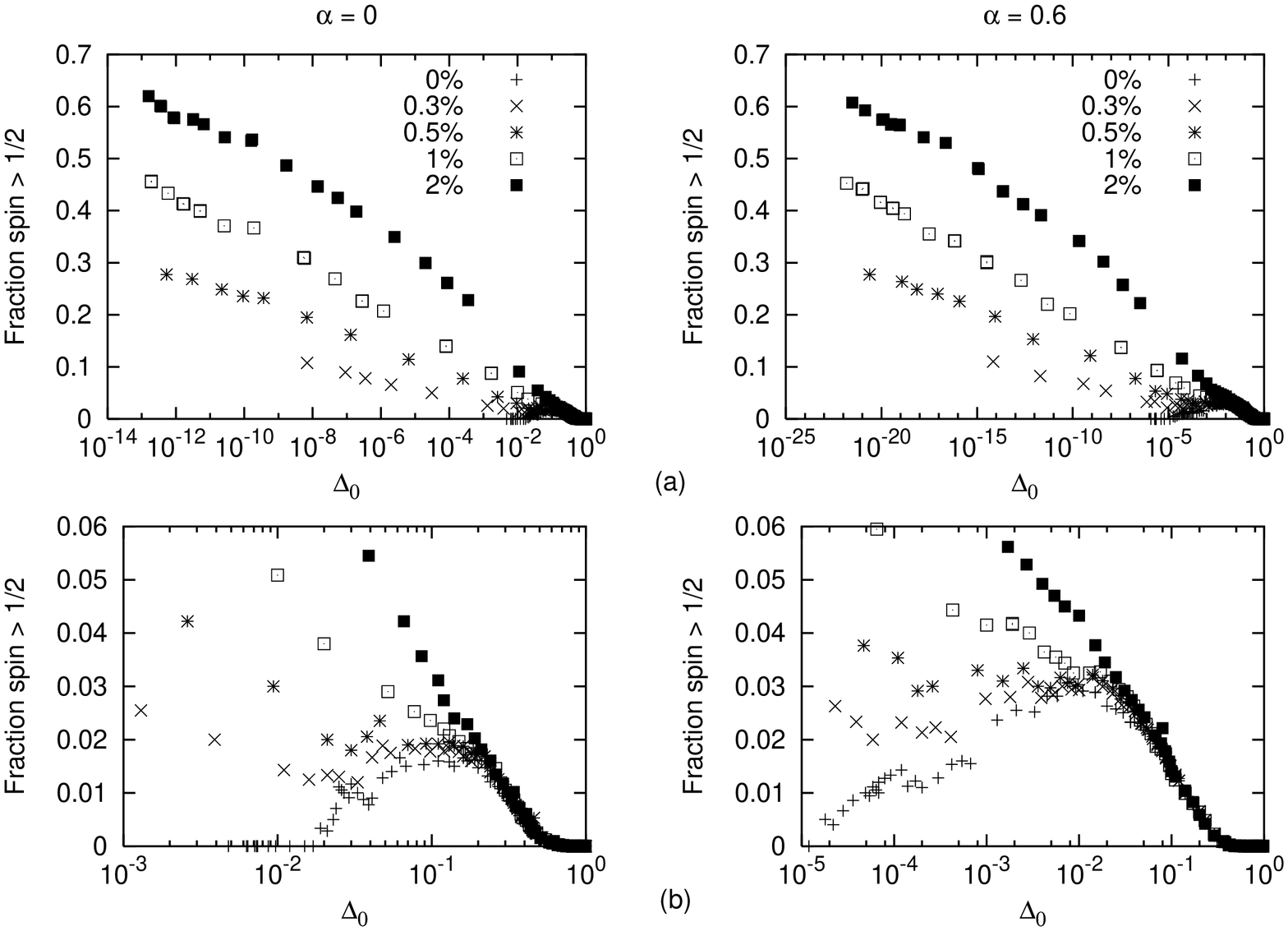}
\includegraphics*[angle=-90,scale=0.6]{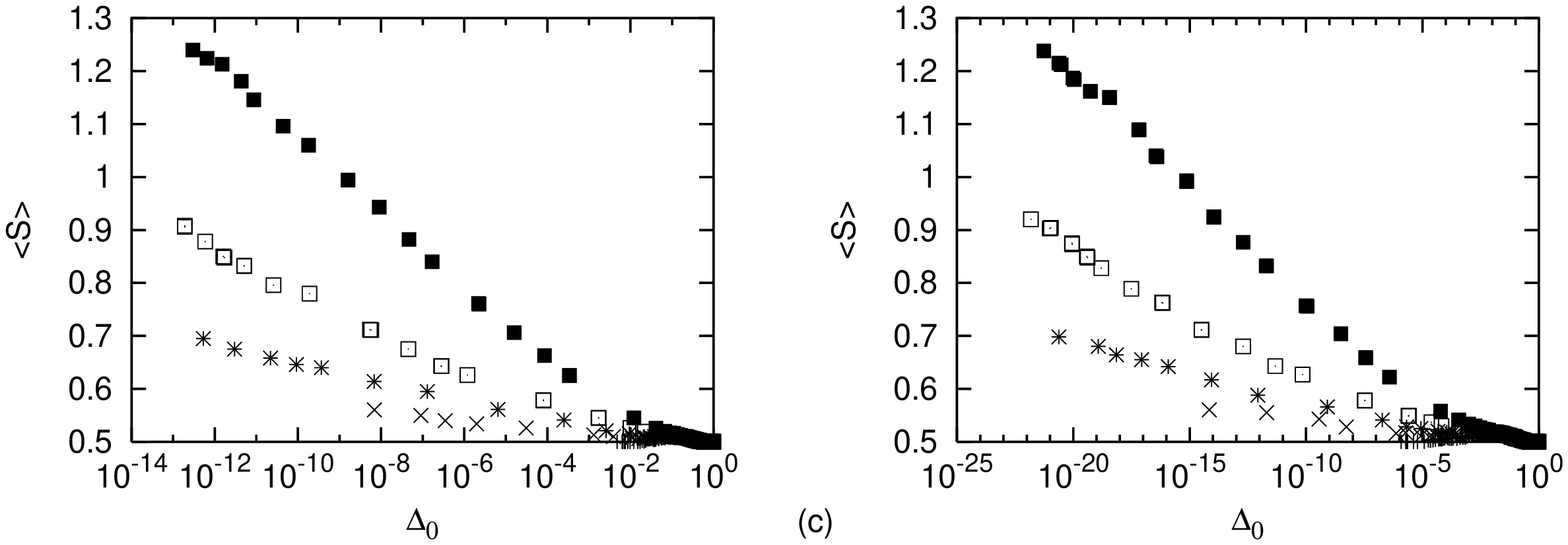}
\caption{The results from numerical calculations for nonmagnetic impurities. 
The left column is for $\alpha = 0$ and the right column for $\alpha = 0.6$, 
both with $\Lambda=1$. The number of spins on a single chain is $N=100 000$.
(a) and (b) The fraction of 
spins larger than 1/2, and (c) the spin size average as a function of cutoff 
$\Delta_0$ with different dopant concentrations. All 
are sample averaged. A more detailed view of the high temperature part from 
(a) is shown in (b).
The error bars, not shown in the figures, are comparable to the size of the
data points.}
\label{fig:spin}
\end{figure*}

We start by discussing the effects of nonmagnetic dopants on the spin ladder.
Due to the generation of F bonds in the RSRG procedure, effective spins 
with sizes bigger than 1/2 are formed as RSRG is carried out. 
The question whether or not these large effective spins 
proliferate at low energy is very important. In the undoped case (with bond
randomness only), we have 
shown \cite{ek} that these large effective spins do {\em not}
proliferate for the reasons discussed in section II.
The situation becomes completely different when 
a finite percentage of dopants are introduced into the ladder. This is shown 
in Fig. \ref{fig:spin}(a) and (b) where we plot the 
fraction of spins larger than 1/2, and in Fig. \ref{fig:spin}(c) where we plot 
the average spin size for different dopant
concentrations as a function of cutoff $\Delta_0$, for $\alpha=0$ and 
$\alpha=0.6$, both with $\Lambda = 1$. For dopant
concentrations bigger than 1\% there is a very clear indication for large spin 
proliferation at low energy. The fraction of spins 
larger than 1/2 grows without bound as the energy scale is lowered. 
This picture is also supported by the results for the average spin size  
which show no sign of decreasing. This is in sharp contrast with zero doping,
which is also included for comparison. 
The behavior for lower dopant
concentrations ($<$ 1\%) is more interesting. 
In this regime we see a clear turnover in 
the graphs where the fraction of spins larger than 1/2 initially increases with
decreasing energy scale, reaches a maximum, then 
it decreases before it begins to rise again (see Fig. \ref{fig:spin}(b)).
Our interpretation of this behavior is the following. 
For very low dopant concentrations, the effects of the dopants are very weak,
and the system behaves like 
an undoped spin ladder at higher energy scales
down to a certain energy scale $\Delta_c$. Below $\Delta_c$, the 
effect of these dopants kicks in and eventually dominates the physics,
and the system flows into the Large Spin 
phase. Thus in a way the dopants are relevant perturbations in the RG sense. 

The proliferation of large effective spins can be understood from the following
analysis.
The nonmagnetic impurities introduced into the system can go 
into either
sublattice A or B with equal probability,
so in \textit{average} the number of dopants in the two 
sublattices are equal. However, due to statistical fluctuations, the number of 
dopants in the two sublattices are {\em not}
equal in {\em specific} realizations of
the random distributions. In particular, in any finite segments of the system,
the fluctuations leave some of the half spins uncompensated for
and destroy the perfect cancellation discussed in Sec. II, and the number of
such spins grows as the square root of the size of the segment. The large 
effective spins that get generated under RG have the same spin size as the 
ground state spin quantum number of the finite segments that they are made of,
thus the proliferation of large spin at low energies is simply reflecting the
growing fluctuation of the spin size of longer and longer segments.

\begin{figure*}
\includegraphics*[angle=-90,scale=0.4]{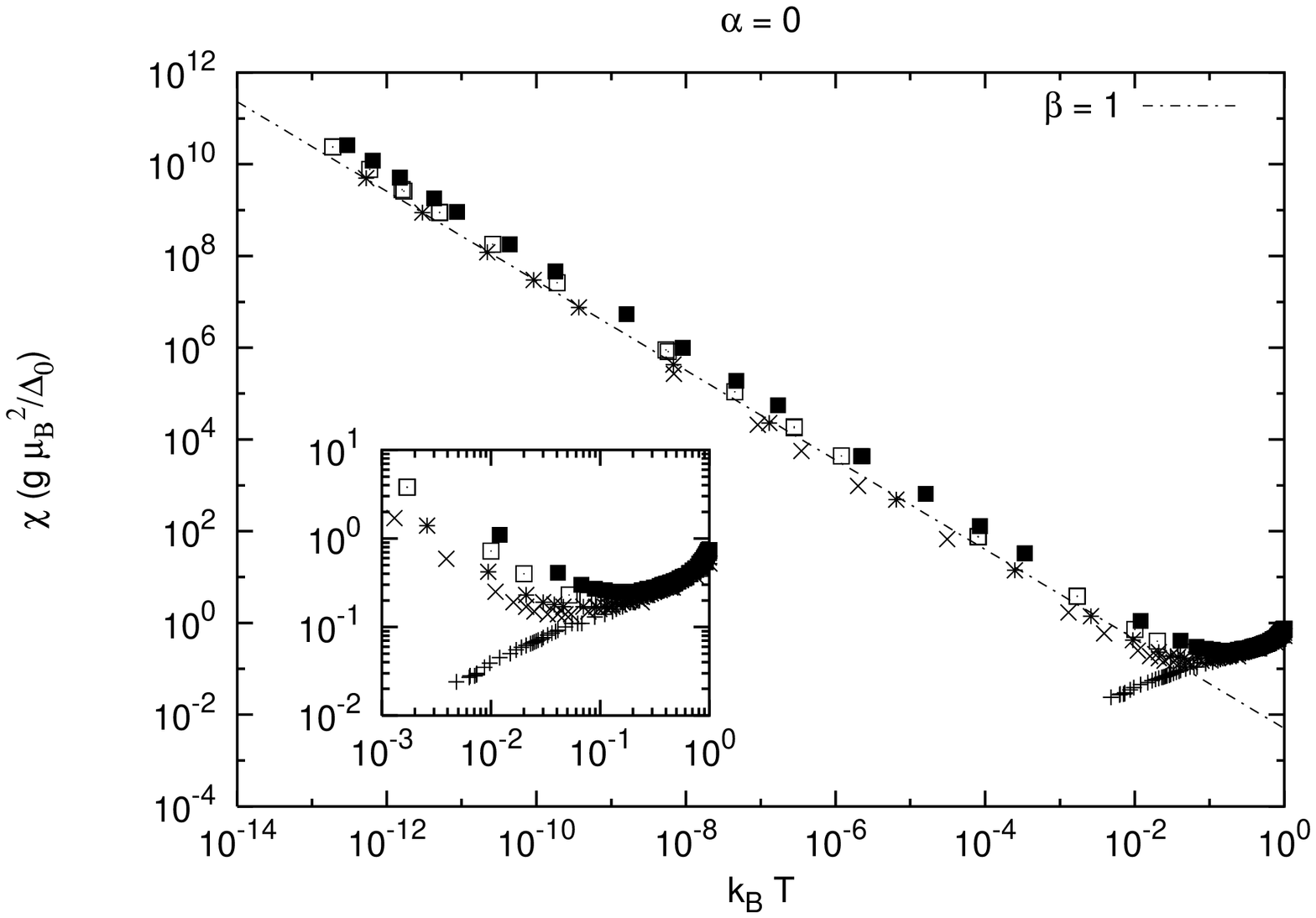}
\includegraphics*[angle=-90,scale=0.4]{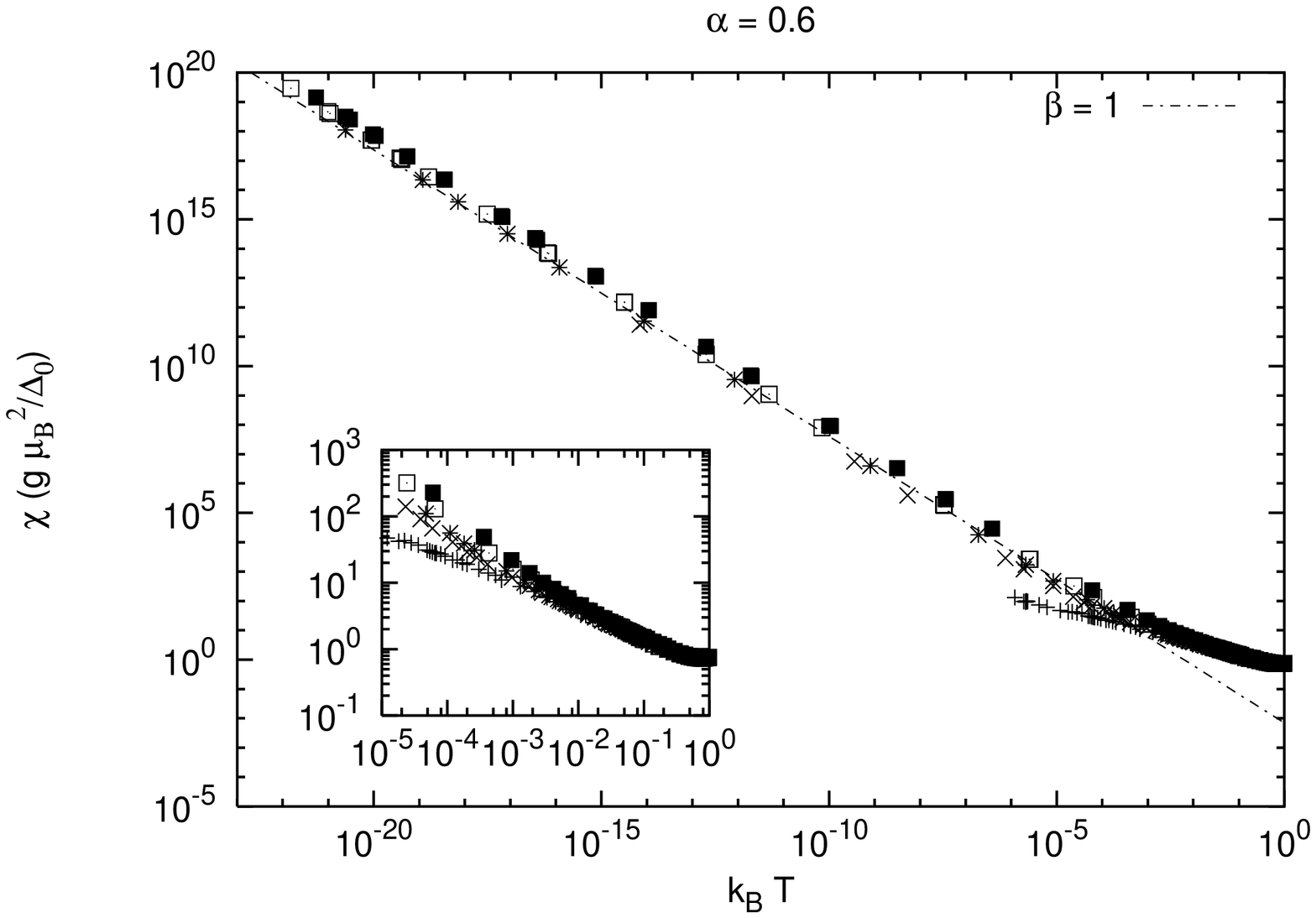}
\caption{The sample averaged susceptibilities per spin as a function of
temperature with different nonmagnetic dopant concentrations for $\alpha=0$
and 0.6. The symbols have the same
meaning as those in Fig. \ref{fig:spin}. The inset shows the part in 
susceptibility that crosses from one behavior to another.
In both cases the $1/T$ line is drawn on the data
as a guide to the eye. 
We do not include the error bars in the figures which are 
comparable to the size of the data points.}
\label{fig:sus}
\end{figure*}

The thermodynamic properties of the doped ladders are dramatically influenced
by the presence of large effective spins in the system.
Fig. \ref{fig:sus} shows the magnetic susceptibility for ladders with
$\alpha=0$ 
and $\alpha=0.6$ with varying dopant concentrations, all with $\Lambda=1$. 
We associate the temperature with the cutoff $\Delta_0$ where we stop the RG
procedure and calculate the contribution from the active spins to the 
susceptibility. These active spins consist of undecimated half spins and 
effective spins larger than 1/2 generated during the decimation process.
All the spins that have been decimated down to the cutoff
$\Delta_0$ do not contribute to the susceptibility. All the active spins
are treated as free spins, so the contribution can be calculated using :
\begin{equation}
\chi_{tot} = \frac{g \mu_B^2}{3k_B T} \sum_s N_s s(s+1),
\end{equation}
where $N_s$ is the number of spins left at energy scale $\Delta_0=k_BT$ for a 
given spin size $s$ and the summation runs over all possible spin sizes.

In all cases we find that at low temperature the susceptibility can be fit 
very well to a Curie-like behavior $\chi\sim T^{-1}$, 
which is {\em insensitive} to specific details of the systems, like the
strength of bond randomness and the dopant concentrations. Such behavior
agrees with the predictions of Sigrist and Furusaki\cite{sigrist} 
but it is very different from what we found earlier on the 
{\em undoped} ladders with bond randomness only,
where the low-$T$ susceptibility follows non-universal power laws with an
exponent that depends on 
the bond disorder strength as well as
the strength of the interchain interactions. We note that while our results
agree with the Curie behavior predicted by Sigrist and Furusaki, due to the
limited temperature range, they may also be fit to the Random-Singlet like
behavior $\chi\sim 1/(T\log^2T)$ that Gogolin {\em et al.}
\cite{steiner,gogolin} suggested, which differs from the Curie 
behavior with a factor that only has a logarithmic dependence on $T$. 
In order to further clarify the
situation, we study the dependence of the Curie coefficient on the parameters 
of the system and compare it with predictions made by Sigrist and Furusaki.

\begin{figure*}
\includegraphics*[angle=-90,scale=0.6]{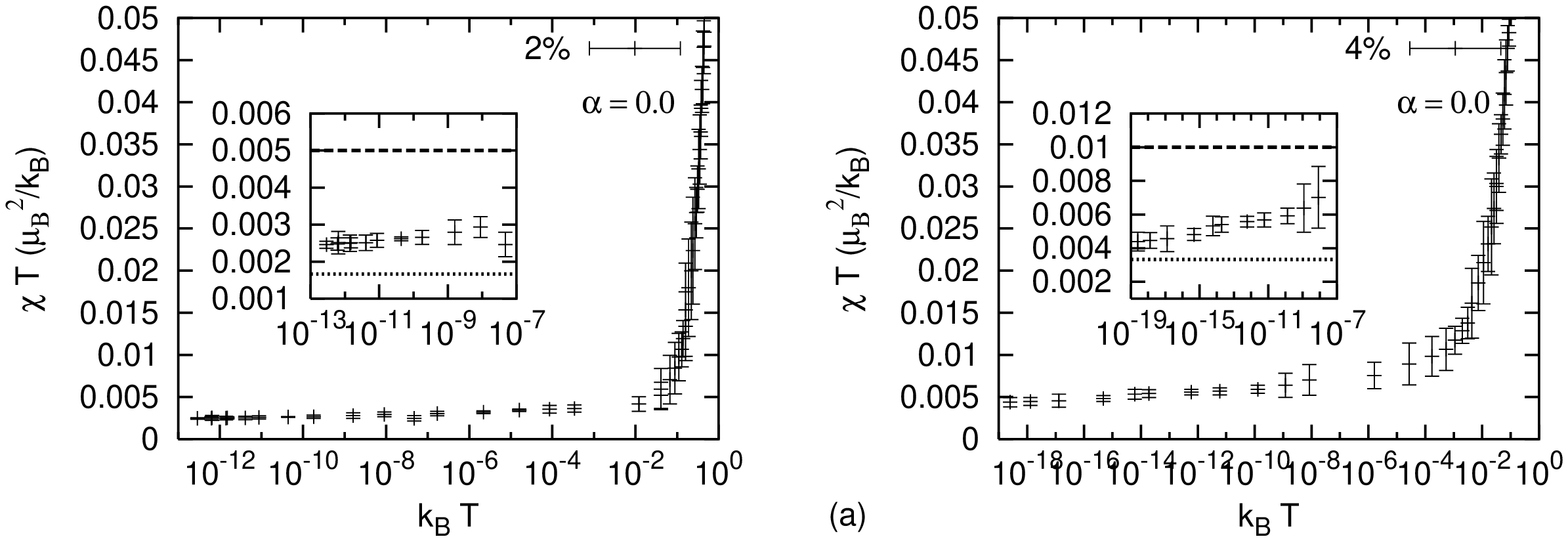}
\includegraphics*[angle=-90,scale=0.6]{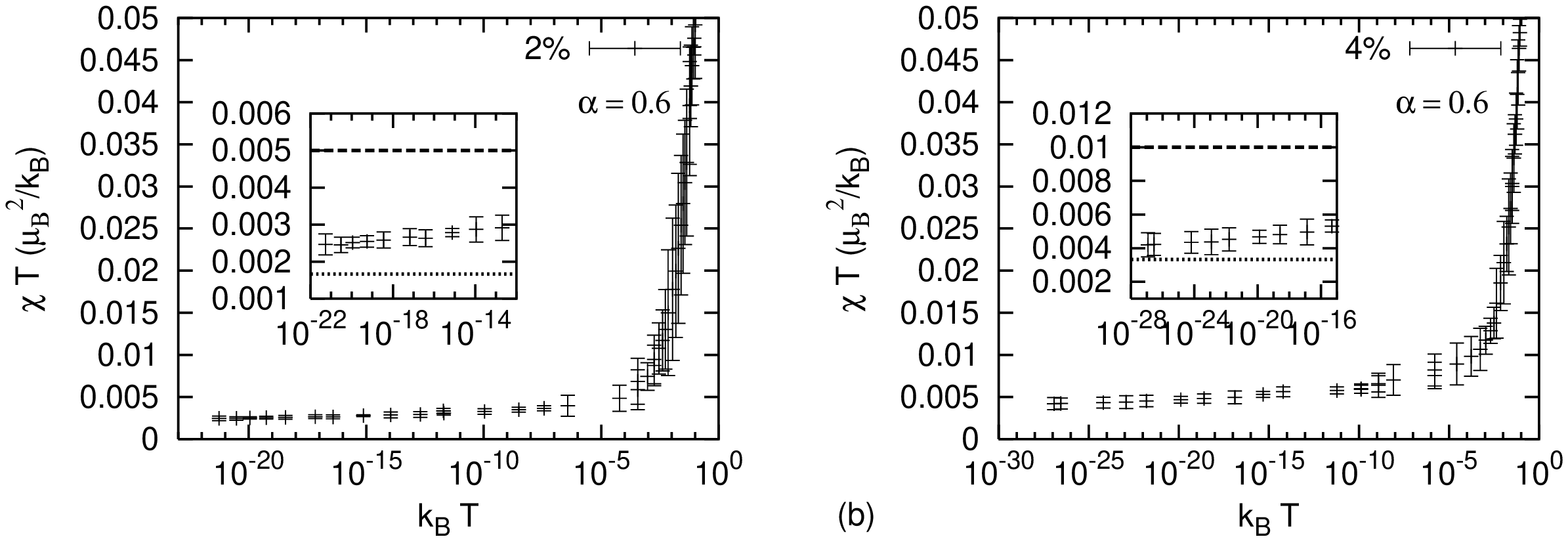}
\caption{The sample averaged Curie constants for (a) $\alpha=0$ and (b)
$\alpha=0.6$. For each $\alpha$ we plot two different dopant concentrations. 
The dashed line
in the inset is the Curie constant for free uncorrelated spin given by 
$z\mu_B^2/4 k_B$ and the dotted line is the constant for strongly correlated
spins given by $z\mu_B^2/12 k_B$. In all cases, the Curie constants are
always approaching the asymptotic limit given by Eq. (\ref{correlate}) in the
low temperature regime.}
\label{fig:curie}
\end{figure*}

The $1/T$ Curie behavior is usually
associated with free spins. In our case however, 
the $1/T$ dependence has a very different origin;
it comes from the strongly correlated effective spins formed during the
RG procedure, due to the existence of ferromagnetic couplings, which form 
clusters whose average size grow in a random walk fashion at low temperature.
Sigrist and Furusaki \cite{sigrist}, in their effective model have shown, 
using the random walk 
argument similar to that used in Ref. \onlinecite{westerberg}, 
that the Curie constant
for finite dopant concentrations is given by :
\begin{equation}
\label{correlate}
\chi T = z\mu_B^2 /(12 k_B).
\end{equation}
On the other hand, if the effective spins induced 
by the dopants behave like free spins, and the Curie constant is given by
\begin{equation}
\label{free}
\chi T = z\mu_B^2 /(4 k_B).
\end{equation}
We plot the Curie constants for $\alpha=0$ and 0.6, each with two different 
dopant concentrations, 2\% and 4\%, as a function of
temperature in Fig. \ref{fig:curie}. The figure shows that at low 
temperature, the Curie constants deviate
significantly from the free spin Curie constant, Eq. \ref{free}, 
and approach the
asymptotic limit, Eq. \ref{correlate}. This strongly suggests that the
effective spins are strongly correlated and the susceptibilities 
follow $1/T$ behavior at low temperature due to large moment formation. 
If the susceptibilities were to follow 
$1/(T$ log$^2T)$,
as Gogolin {\em et al.}\cite{gogolin} suggested, the Curie constants would go 
to zero at low temperature. While we do see that the Curie constants 
decrease with decreasing $T$,  
they are approaching constants given by Eq. \ref{correlate} in the low-$T$
limit, instead of going to zero. We thus conclude that our data strongly
support the results of Sigrist and Furusaki.\cite{sigrist}
We note that
Miyazaki {\em et al.}\cite{miyazaki} used quantum Monte Carlo method to 
calculate the Curie coefficients of the doped ladder with different dopant
concentrations. They were unable to
obtain conclusive results for the coefficients
due to the fact that the system size studied 
was not large enough to probe deep into the low temperature regime.

The temperature dependence of susceptibility also 
gives us some information how the system 
crosses over from one behavior (at high $T$) to another (at low $T$). 
In Fig. \ref{fig:sus}
we plot the susceptibility with different dopant concentrations for $\alpha=0$
and $\alpha=0.6$ as a function of temperature. The inset of each figure 
shows the part of susceptibilities where the crossover into a new
behavior occurs.
This crossover is particularly clear for $\alpha=0$. 
As we vary the dopant concentrations, from 0\% to 2\%, there is 
a clear turnover in the susceptibilities. In the undoped limit, the 
susceptibility goes to zero as $T\rightarrow 0$. 
For very small $z$, $\chi$ follows this behavior at higher $T$, as the effect 
of the dopants have not yet dominated the contribution to $\chi$.
However at low enough temperature the effects of the dopants start to dominate;
this is characterized as the susceptibility begins to increase and
finally becomes divergent as
the temperature is decreased below a certain crossover scale. 
The same behavior can also be
seen for $\alpha=0.6$ although it is not as pronounced as for $\alpha=0$, 
because in the undoped limit, $\chi$ is already divergent as a power-law of 
$T$, and the power law exponent is given by $\beta \approx 0.4$.
\cite{ek} Introducing a small amount of dopants into the system alters
the physics at sufficiently low temperature where the susceptibilities have
different power law exponent. 

One important observation from the numerical results is that in the presence of
a finite dopant concentration, the physics of the 
systems in the low temperature limit is {\em not}
sensitive to the choice of the distribution of the bond randomness.
The low energy physics for two quite different bond randomness 
strength $\alpha=0$ and $0.6$, as shown in Fig. \ref{fig:spin}, \ref{fig:sus}, 
and \ref{fig:curie}, are essentially the same. 
In both cases the systems are controlled by large 
effective spins at low energy and the susceptibilities follow $1/T$ behavior
at low temperature,
and the Curie constants are approaching the same asymptotic limit,
given by Eq. \ref{correlate}, which depends on dopant concentration 
{\em only}. 
Thus the insensitivity of the results on the specific form of the bond
distribution justifies our choice of the bond distribution based on 
convenience.
We note that it tends to flow to a power-law form even if it does not have
such form initially; thus by choosing such a form it puts one closer to the
asymptotic form and reduces finite size effects. 

\begin{figure*}
\includegraphics*[angle=-90,scale=0.6]{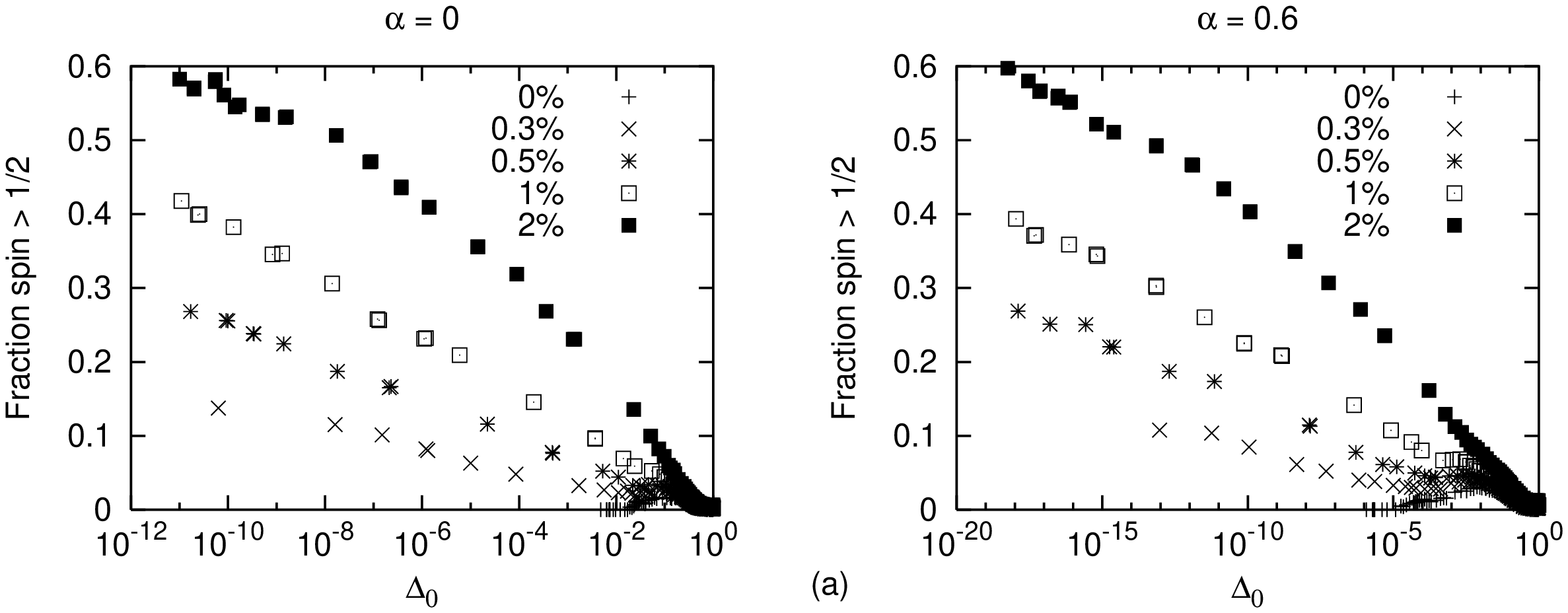}
\includegraphics*[angle=-90,scale=0.6]{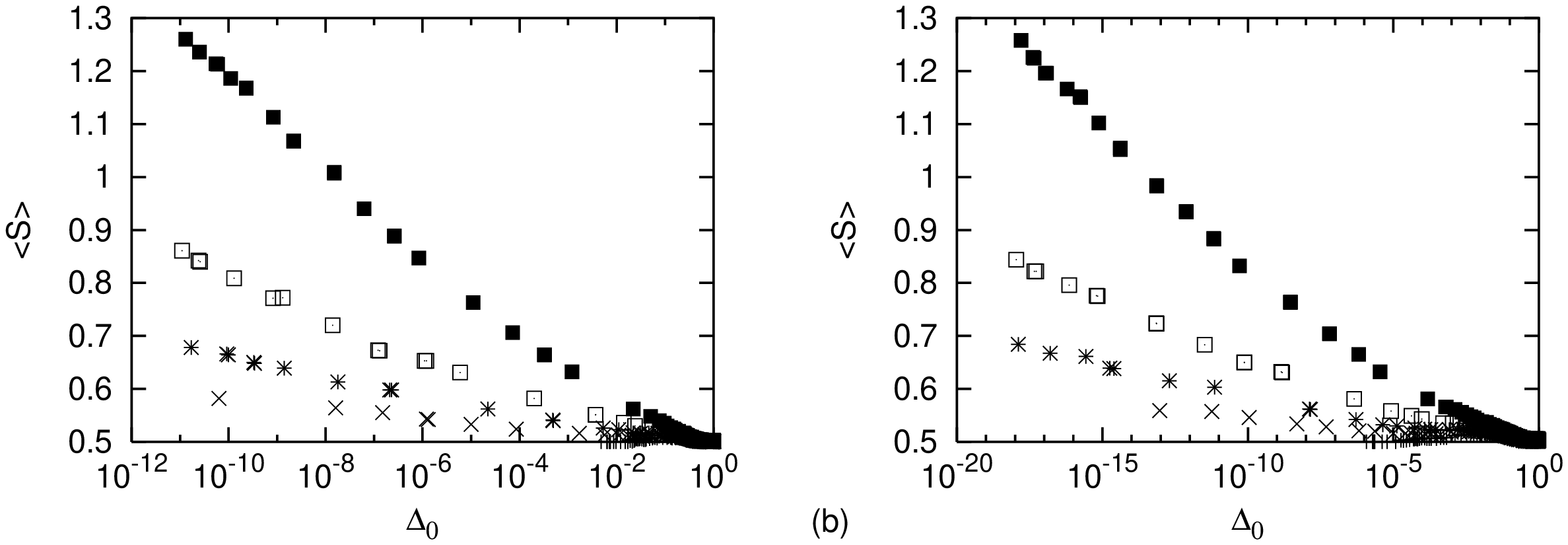}
\includegraphics*[angle=-90,scale=0.6]{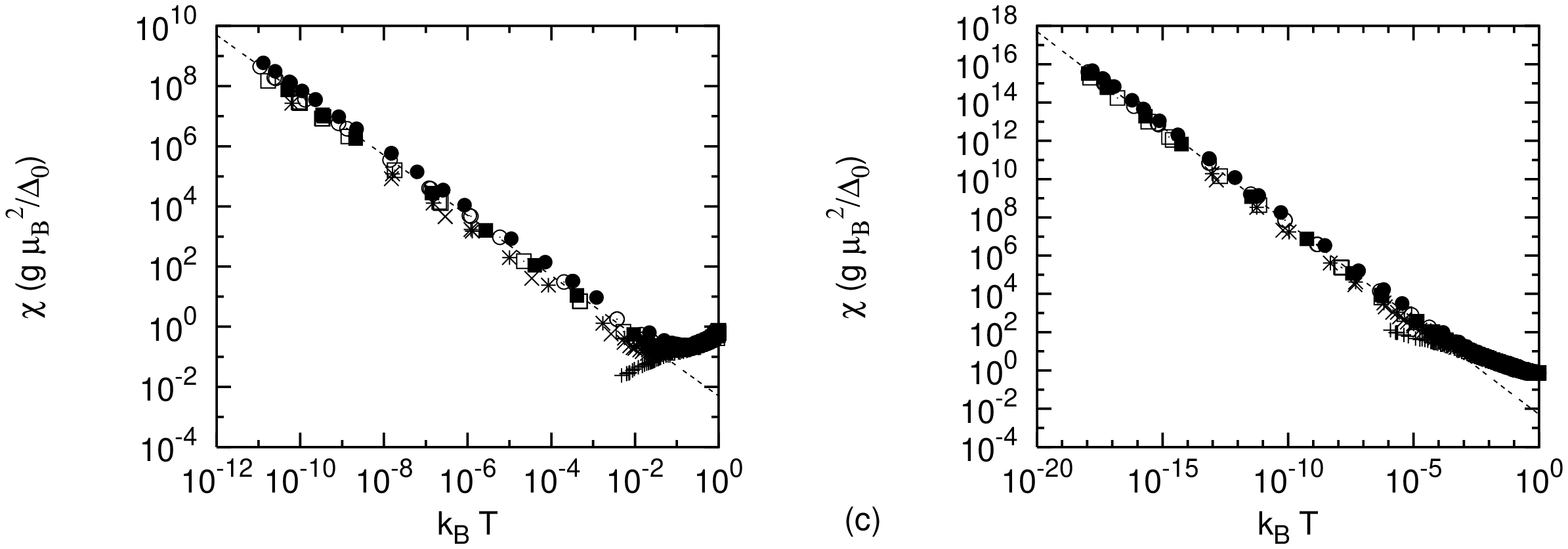}
\caption{The results from numerical calculations for magnetic impurities. 
The left column is for $\alpha = 0$ and the right column for $\alpha = 0.6$, 
both with $\Lambda=1$. The number of spins on a single chain is $N=100 000$.
(a) The fraction of 
spins larger than 1/2, (b) the average spin size as a function of cutoff 
$\Delta_0$ with different dopant concentrations, and (c) the sample averaged 
susceptibilities per spin as a function of temperature with different 
magnetic dopant concentrations. The dashed line in (c) is the $1/T$ line drawn
as a guide to the eye.}
\label{fig:magnetic}
\end{figure*}

We now turn our discussion to the effects of magnetic dopants with spin-1
on the spin ladders, which turn out to be very similar to those of
non-magnetic 
dopants.
In Fig. \ref{fig:magnetic} we plot 
the fraction of spin larger than 1/2, the average spin size as a function of 
cutoff $\Delta_0$, and the susceptibility as a function of 
temperature with different magnetic dopant concentrations.
As we can see from these figures, the qualitative 
behavior of the system doped with magnetic impurities at low energy are the 
same as that doped with nonmagnetic impurities. Large spin formations 
are seen at low energies which grow continuously as the energy is
decreased. The similarity in the effects 
of these two different types of dopants lies in the 
fact that they both induce spin-1/2 local moments, and
uncompensated spins in finite segments, due to the
fluctuation in the number of dopants going into the two different sublattices.
This is the origin of the proliferation of large effective spins at low 
energies, and the $1/T$ Curie dependence of the susceptibility.

\section{Summary and Discussions}

We have studied antiferromagnetic two-leg spin-1/2 ladders with bond
randomness and site dilution/magnetic impurity
by means of a real-space renormalization-group
scheme.
We found that there is proliferation of large effective spins at low energy for
non zero dopant concentrations. These large effective spins show the tendency 
of growing without bound as the energy scale is lowered. 
The susceptibility of the doped spin ladder follows Curie-like $1/T$ behavior
at low temperature. This behavior remains universal regardless of the dopant
concentrations and the strength of bond randomness.
We also find the Curie coefficient is controled
by the dopant concentration only, and agrees with
that predicted by 
Sigrist and Furusaki.\cite{sigrist} 
We conclude that non zero dopant concentrations always drive the system into a
phase dominated by the large effective spins. The large effective spins
control the low temperature physics of the system which makes the the doped
ladder behaves in many respects like the random spin chain with random
ferromagnetic
and antiferromagnetic interactions.
This is very different from what we found in our
earlier work for ladders with bond randomness only,\cite{ek} where no
large spin proliferation was found, and
the susceptibility at low temperature
follows non-universal pwoer law: $\chi\sim T^{-\beta}$ with an 
exponent $\beta$ that depends on the 
strength of bond randomness and the strength of interchain interactions.

Unfortunately at present we cannot make a direct comparison between our results
and experiments,
because in the systems studied so far the doped ladders
all form long-range antiferromagnetic order at low temperature, due to the 
presence of 3D inter-ladder couplings not included in our study.

While reaching the same conclusions, we used a different approach in our
study of the doped ladders as compared with the work of
Sigrist and Furusaki.\cite{sigrist} They focused exclusively on 
the {\em effective} spins that are induced by the dopants, and neglected all 
the original spins, justified by the fact that without the dopants and the
effective spins they induce, the system is gapped. Thus the model they used is
an {\em effective} model appropriate for describing the low-$T$ properties of
the system. In our study on the other hand we include all the original degrees
of freedom (the original spins), and systematically lower the energy scale
by decimating strong bonds one by one. Our approach thus treats high- and 
low-energy degrees of freedom on equal footing, and allows us to address both
the high-$T$ and low-$T$ properties of the systems, and the crossover between 
them. Thus our study is complementary to that of Sigrist and Furusaki.

\acknowledgments
This work was supported by NSF grants No. DMR-9971541 and DMR-0225698, 
the Research 
Corporation, and the Center for Materials Research and Technology (MARTECH). 
K.Y. was also supported in part by the A. P. Sloan Foundation.

\begin{thebibliography}{99}
\bibitem{note} The specific distributions of the $J$'s in real systems are not
known. As we will see however, our results are not sensitive to the choice of
these distributions; see later. So we may choose the distributions based on
convenience of the study. Also from previous studies it is known that the
distributions tend to flow to power-law forms at low energies, even if they do
not take such forms initially.
\end{thebibliography}
\end{document}